\documentclass[useAMS,usenatbib,usegraphicx]{mn2e}

\usepackage{amsmath}
\usepackage{breqn}
\newcommand{\msun}{M$_{\odot}$}
\newcommand{\ms}{$M$--$\sigma$ }

\def\apj{\rm ApJ}
\def\apjl{\rm ApJL}
\def\apjs{\rm ApJS}
\def\aj{\rm AJ}
\def\mnras{\rm MNRAS}
\def\nat{\rm Nature}

\def\araa{\rm ARAA}

\title[Inclination Effects on Velocity Dispersion]{Effects of
  Inclination on Measuring Velocity Dispersion and Implications for
  Black Holes} \author[Bellovary et al.]{Jillian
  M. Bellovary$^{1,2}$\thanks{E-mail:
    jillian.bellovary@vanderbilt.edu} ,Kelly Holley-Bockelmann$^{1,2}$,Kayhan G{\"u}ltekin$^{3}$,   Charlotte \newauthor R. Christensen$^{4,5}$, Fabio Governato$^{6}$, Alyson M. Brooks$^{7}$, Sarah Loebman$^{3}$,   Ferah Munshi$^{8}$\\
  $^{1}$Department of Physics and Astronomy, Vanderbilt University, PMB 401807, Nashville, TN, 37206, USA\\
  $^{2}$Department of Natural Sciences and Mathematics, Fisk University, 1000 17th Avenue N., Nashville, TN 37208, USA\\
  $^{3}$Department of Astronomy, University of Michigan, 830 Dennison, 500 Church St., Ann Arbor, MI, 48109, USA\\
  $^{4}$Department of Astronomy, University of Arizona, 933 North Cherry Avenue, Rm. N204, Tucson, AZ, 85721, USA\\
   $^{5}$Department of Physics, Grinnell College, 1116 Eighth Ave, Grinnell, IA, 50112, USA\\
  $^{6}$Department of Astronomy, University of Washington, Box 351580, Seattle, WA, 98195, USA\\
  $^{7}$Department of Physics and Astronomy, Rutgers University, 136 Frelinghuysen Rd, Piscataway, NJ, 08854, USA\\
  $^{8}$Department of Physics and Astronomy, University of Oklahoma, 440 W. Brooks St.,  Norman, OK, 73019, USA}

\begin{document}


\pagerange{\pageref{firstpage}--\pageref{lastpage}} \pubyear{2002}

\maketitle

\label{firstpage}

\begin{abstract}
  The relation of central black hole mass and stellar spheroid
  velocity dispersion (the \ms relation) is one of the best-known and
  tightest correlations linking black holes and their host galaxies.
  There has been much scrutiny concerning the difficulty of obtaining
  accurate black hole measurements, and rightly so; however, it has
  been taken for granted that measurements of velocity dispersion are
  essentially straightforward.  We examine five disk galaxies from
  cosmological SPH simulations and find that line-of-sight effects due
  to galaxy orientation can affect the measured $\sigma_{\rm los}$ by
  30\%, and consequently black hole mass predictions by up to 1.0 dex.
  Face-on orientations correspond to systematically lower velocity
  dispersion measurements, while more edge-on orientations give higher
  velocity dispersions, due to contamination by disk stars when
  measuring line of sight quantities.  We caution observers that the
  uncertainty of velocity dispersion measurements is at least 20 km
  s$^{-1}$, and can be much larger for moderate inclinations.  This
  effect may account for some of the scatter in the locally measured
  \ms relation, particularly at the low-mass end.  We provide a method
  for correcting observed $\sigma_{\rm los}$ values for inclination
  effects based on observable quantities.
\end{abstract}

\begin{keywords}
galaxies: bulges, galaxies: spiral, galaxies: kinematics and dynamics, methods: numerical
\end{keywords}

\section{Introduction}

One of the most critical discoveries in recent years is the apparent
co-evolution of central supermassive black holes (SMBHs) and their
host galaxy spheroids.  This phenomenon, often represented in scaling
relations such as \ms, $M_{BH} - M_{bulge}$, or $M_{BH}-L_{bulge}$,
has been observed to hold over several orders of magnitude of SMBH
mass and a variety of galaxy properties
\citep[e.g.][]{Magorrian98,Ferrarese00,Gebhardt00,Tremaine02,Marconi03,Haring04,Gultekin09,Graham11,McConnell13}.
Further observational campaigns suggest that these relations may
evolve with redshift \citep{Peng06,Treu07,Woo08,Decarli10,Bennert11},
though others refute this claim
\citep{Lauer07b,VolonteriStark11,Schulze14}.  The \ms relation is a
key constraint on any theory of the interplay between SMBH growth and
galaxy evolution~\citep[e.g.][]{Loeb94, Haehnelt00,
  Granato01,Menou01,DiMatteo05, Wyithe05, Croton06,
  Micic07,Hopkins08,Tanaka09,Volonteri09,Micic11,
  Bellovary13,KormendyHo}. This relation is so well-accepted that both
theoretical and observational studies use the \ms fit to scale the
SMBH mass within a galaxy when not directly observable
\citep{Volonteri03,Somerville08,Wild10}. To build a theory of SMBH
assembly in the context of galaxy evolution, it is clear that we need
both accurate measurements of SMBH mass and bulge velocity dispersion,
and a deep understanding of the biases and limits of these
measurements.

Much scrutiny has been given to the difficulty of measuring SMBH
masses, and for good reason; accurate mass measurements are very
difficult and require a large investment of observational resources
and careful analysis.  However, measuring the velocity dispersion,
$\sigma$, of a galaxy spheroid is also non-trivial.  The galaxy
orientation is imprinted on any observational measurement of $\sigma$,
and unless we understand how this effect biases $\sigma$, we are at
the mercy of the structure and viewing angle of every galaxy we
observe. Velocity dispersions are commonly measured spectroscopically
via the widths of stellar absorption lines, but it is difficult to
isolate the light from spheroid stars from those of the disk.  Every
measurement of $\sigma$ of the spheroid, therefore, will be
contaminated by the kinematics of other galaxy components.

One way to examine the effect of orientation on measurements of
$\sigma$ is through simulations.  A simulated galaxy can be rotated
and viewed at any orientation, and can be analyzed to determine the
intrinsic galaxy properties with no observational biases.  We employ a
sample of disk-dominated galaxies and examine how the viewing angle
affects the apparent central velocity dispersion.  We choose disk
galaxies because the effects of orientation will be the most severe,
and we wish to investigate the repercussions for the low-mass end of
the \ms relation, which exhibits relatively large scatter.  The
scatter has been postulated to be due to evolutionary effects, such as
merger history and environment \citep{Kormendy11,Micic11,Mathur12},
and is dependent on galaxy mass, morphology, and bulge/disk ratio,
among other things \citep{Hu08,Graham09}.  However, another possibility is
that some (or all) of the scatter is actually caused by orientation
effects \citep{Gebhardt00}, which include line-of-sight contamination
from disk and halo stars as well as non-symmetric bulge effects and
bulge rotation.  It is thus critical to quantify the effect of viewing
angle when measuring $\sigma$; our understanding of how SMBHs and
galaxies grow depends on it.

In this paper, we examine five disk-dominated simulated galaxies with
a range of masses and bulge sizes.  These galaxies are selected from
``zoomed-in" cosmological simulations and have realistic star
formation histories, baryon and gas fractions, and bulge and disk
scale lengths.  In $\S$ \ref{sect:sims} we describe these simulations
in detail, along with our methodology for measuring $\sigma$.  In
Section 3 we present our results and provide a correction factor for
observed values of $\sigma$.  In Sections 4 and 5 we discuss the
repercussions for the observed \ms relation and summarise the work.

\section[]{Simulations and Velocity Dispersion Measurements}\label{sect:sims}

We use the $N$-Body Smoothed Particle Hydrodynamics code GASOLINE
\citep{Stadel01,Wadsley04} to create ``zoomed-in'' cosmological
simulations of disk galaxies with a range of masses.  A cosmological
context is critical for this study, since it is important that galaxy
bulges build naturally without assumptions as to the kinematics of the
bulge or disk stars.  We select our galaxies from a uniform, dark
matter only 50 comoving Mpc box, and resimulate them using the volume
renormalization method of \citet{Katz93} to better resolve our regions
of interest.  In a box of this size, the fundamental mode is
nonlinear; while this effect changes the number and structure of the
most massive halos, it has negligible effect on the centres of Milky
Way galaxies.  Our gas, dark, and star particle masses are $m_{gas} =
2.7 \times 10^4$ \msun, $m_{dark} = 1.3 \times 10^5$ \msun, and
$m_{star} = 8.0 \times 10^3$ \msun, respectively, and the force
resolution is 174 pc.  The initial conditions were generated with a
WMAP 3 cosmology \citep{WMAP3} and were run from $z = 150$ to $z = 0$.
At $z=9$, a uniform ionizing UV background appears, following the
model of \citet{Haardt01}.  We identify individual galaxies using the
tool AHF \citep{Knollmann09,Gill04}, which finds spherical
overdensities with respect to the critical density.  Changing the
cosmology and reionization technique may affect the total luminosity
function and formation time of low-mass halos, but has little effect
on the spheroid velocity dispersion or bulge-to-disk ratios of our
selected galaxies.

Gas cooling occurs via metal lines \citep[described in][]{Shen10} and
H$_2$ \citep{Christensen12}.  This low-temperature cooling, in
combination with H$_2$ self-shielding and the dust shielding of HI and
H$_2$, allows gas to reach the high densities ($\rho \sim 100$ amu
cm$^{-3}$) and low temperatures ($\la 1000$K) needed to
appropriately model star formation in cosmological simulations
\citep{Governato12}.  Star formation is dependent on the H$_2$
fraction (which itself depends on metallicity and the self-shielding
ability of the gas).  Star particles are born with a Kroupa IMF
\citep{Kroupa}, which dictates the occurrence of supernovae.  Each
supernova deposits $10^{51}$ erg of energy into the ISM within a blast
radius described by \citet{McKee77}.  The gas particles within the
blast radius have their cooling ability quenched until such time as
the blastwave equations allow.  This process mimics a supernova
remnant through the snowplow phase and is described in detail in
\citet{Stinson06}.  We anticipate minimal effect of the choice of IMF
or supernova feedback prescription on the structure and kinematics of
the bulge in these galaxies.  We do not include supermassive black
hole physics in these simulations; we discuss the repercussions in
Section \ref{sect:caveats}.  In short, while activity from
supermassive black holes is expected to modify the central regions of
galaxies, such phenomena are more pronounced among galaxies more
massive than $L_*$ \citep{Fanidakis13}, and we do not expect a
significant effect here.

\begin{figure*}
\includegraphics[scale=0.35]{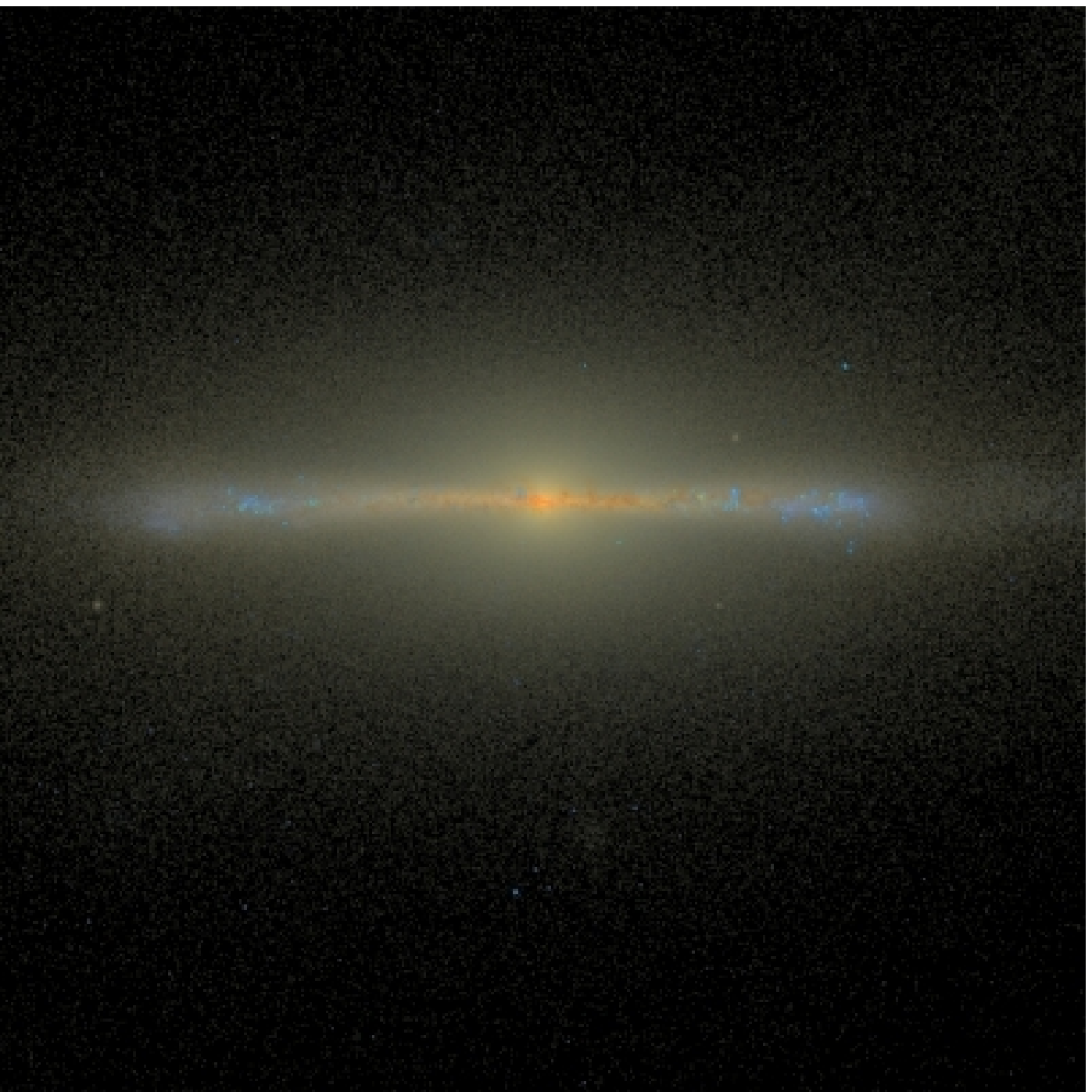}
\includegraphics[scale=0.35]{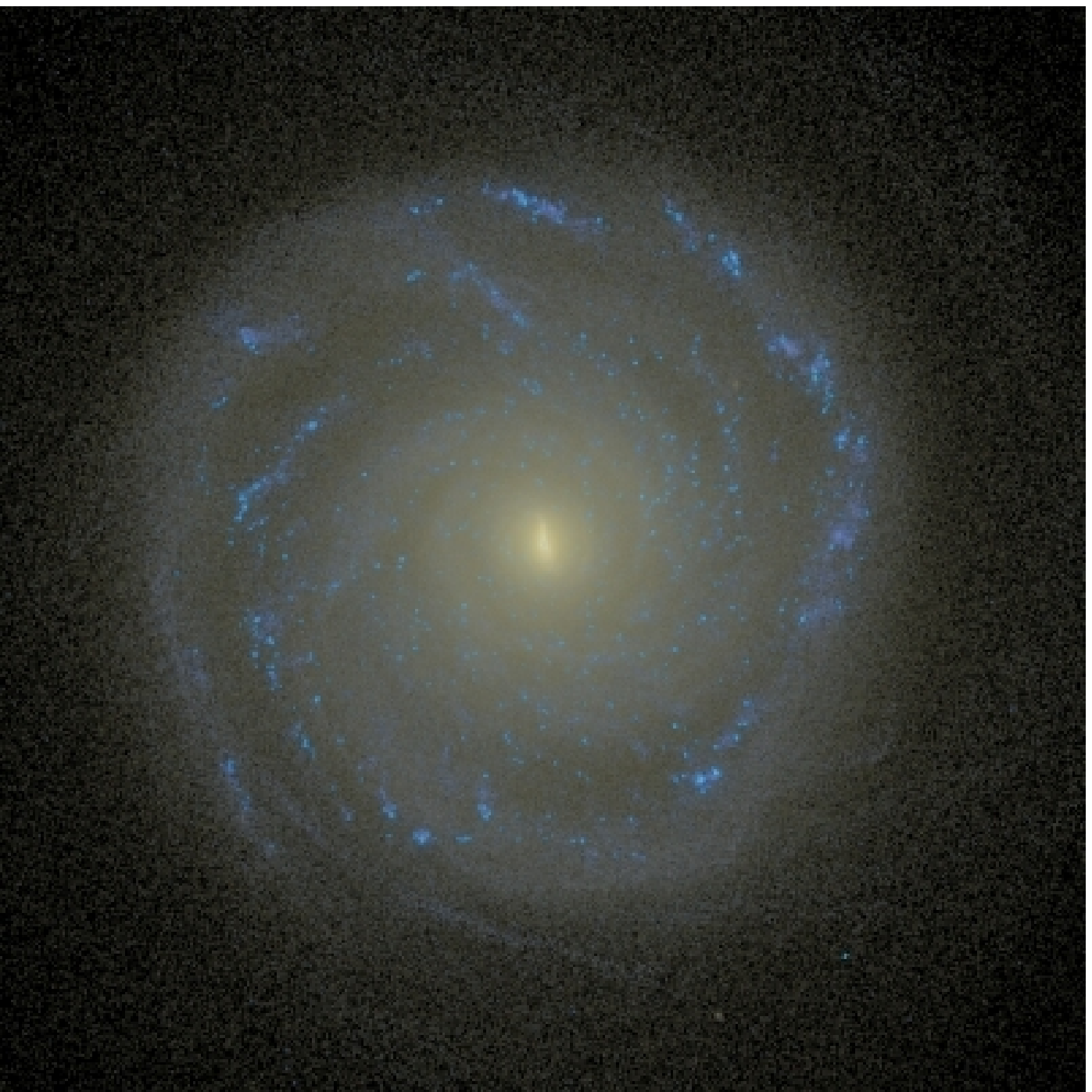}
\includegraphics[scale=0.35]{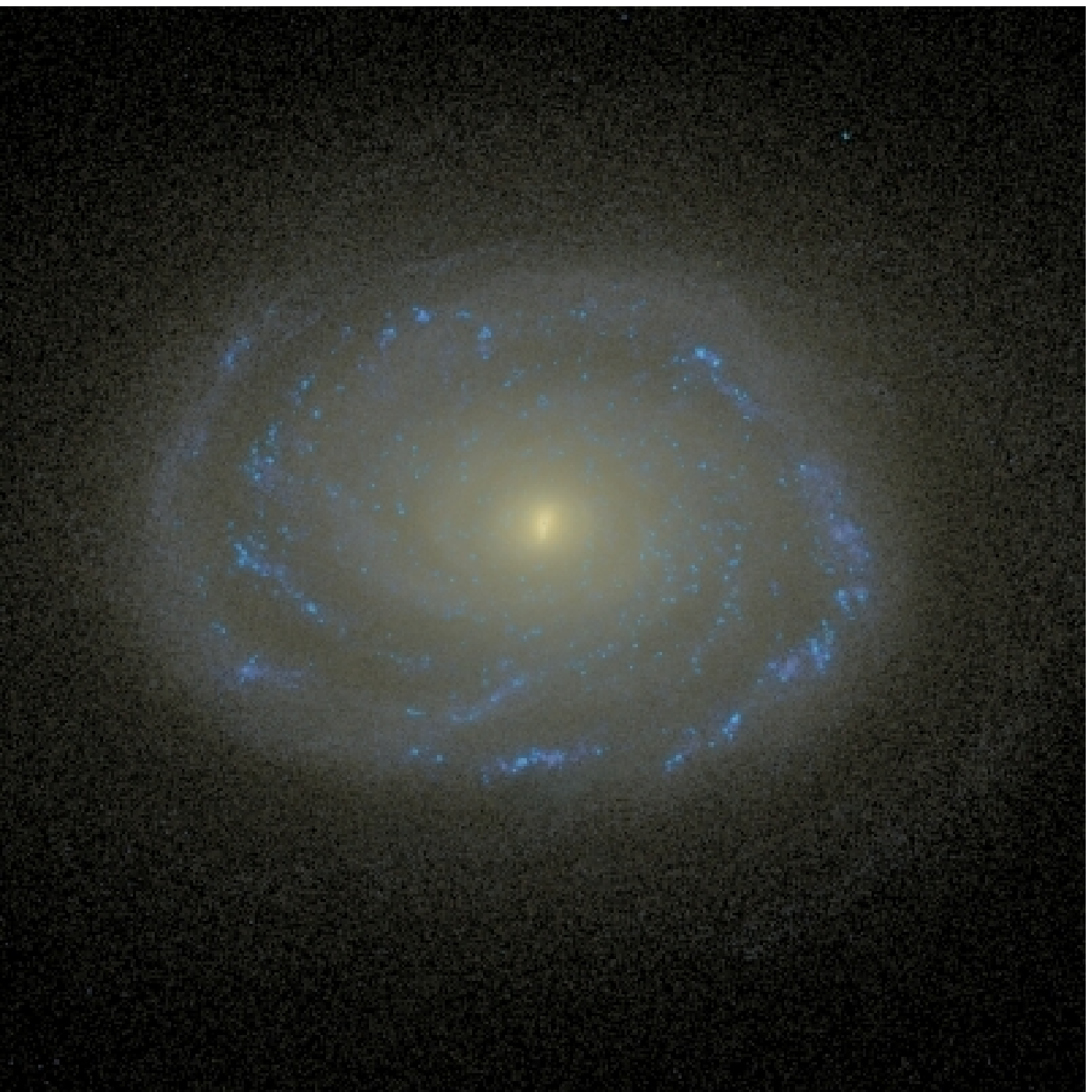}
\includegraphics[scale=0.35]{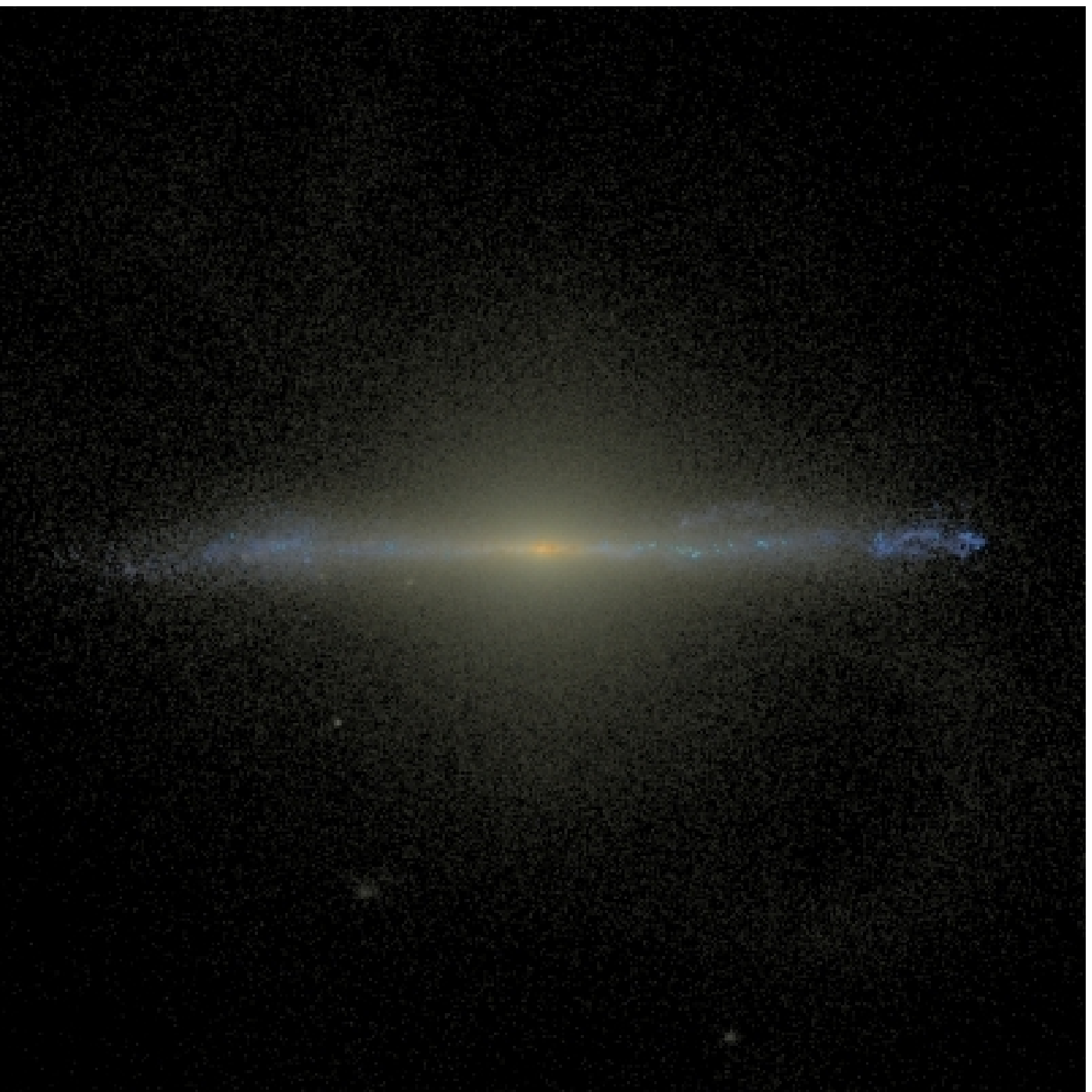}
\includegraphics[scale=0.35]{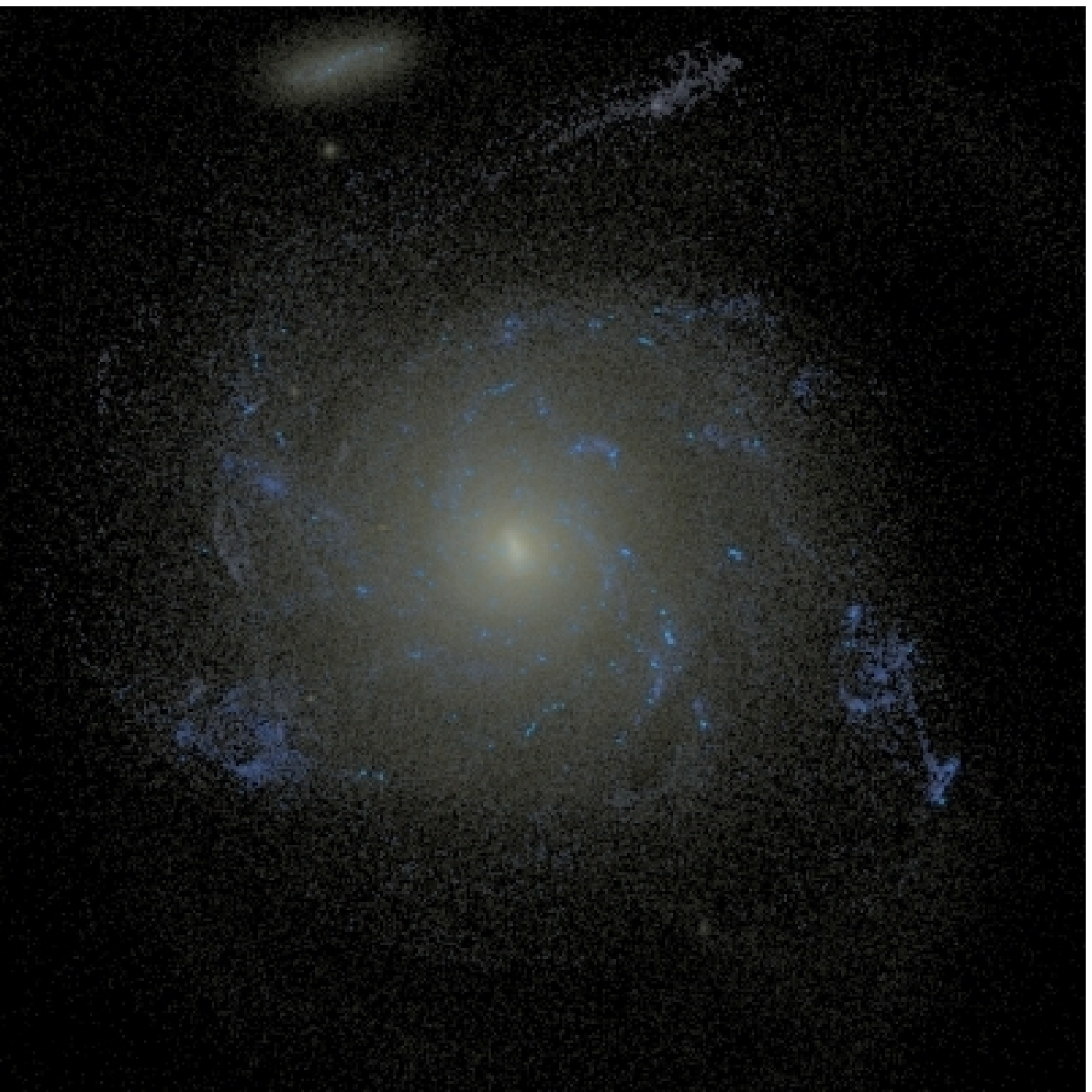}
\includegraphics[scale=0.35]{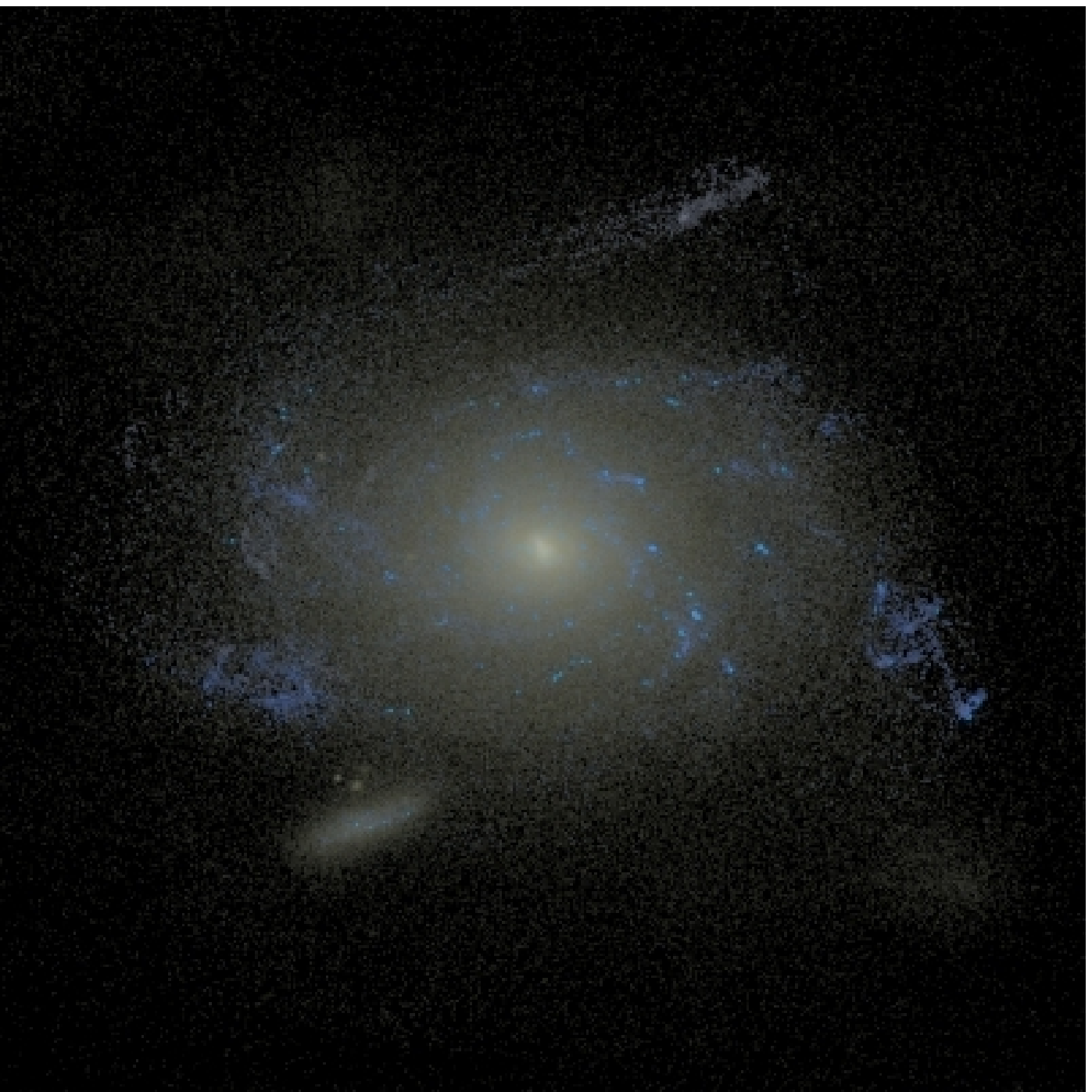}

 \caption{{\em Top row, Left:} Edge-on SDSS $gri$ image of Milky Way-like galaxy $h258$, created with Sunrise \citep{Jonsson06}.  {\em Centre:} Face-on image.  {\em Right:} Image of galaxy   inclined at 45 degrees.  {\em Bottom row:} The same for less-massive galaxy $h603$.  The size of each image is 40 kpc across.
   \label{fig:sunrise}}
\end{figure*}

From each simulation we use the primary galaxy at redshift $z = 0$,
whose properties are detailed in Table \ref{table:properties}.  Our
sample spans a range of 3 - 9 $\times 10^{11}$ \msun~ in total mass
and each galaxy has a prominent bulge and disk (see Figure
\ref{fig:sunrise} for examples).  We focus on disk galaxies because
elliptical galaxies will have less variation in their velocity
dispersion measurements due to orientation effects, and we are
specifically interested in the low mass end of the \ms relation.  All
of the systems we study are relaxed at $z = 0$ (for an interesting
analysis of measuring $\sigma$ in merging galaxies see
\citet{Stickley14}).  The five simulated galaxies for which we measure
bulge velocity dispersion have made previous appearances in the
literature in
\citet{Zolotov12,Loebman12,Christensen14a,Munshi13,Christensen14b}.
GASOLINE has proven to simulate galaxies with realistic baryon
fractions and stellar masses for their halo mass \citep{Munshi13},
bulge and disk properties
\citep{Brooks11,Christensen14a,Christensen14b}, satellite distributions
\citep{Zolotov12,Brooks14}, and which follow the observed
Kennicutt-Schmidt relation \citep{Christensen12}.  One of our
simulations, $h603$, was previously shown by \citep{Christensen14a} to
lie along the bulge scaling relations, demonstrating that the bulge
has the appropriate size and surface brightness with respect to its
host galaxy.  Four of the five galaxies have classical bulges, as
defined by having a Sersi{\'c} index $n > 2$; the galaxy $h603$ is
reported to have $n = 1.65$ in \citet{Christensen14a} and can be
classified as a pseudobulge.  In summary, we are confident that our
simulations realistically represent galaxy bulges and disks, and offer
an excellent setting to explore orientation effects on bulge
properties.


\begin{table}
 \centering
  \caption{Simulation Properties\label{table:properties}}
  \begin{tabular}{@{}lcccllc@{}}
  \hline
   Run     & N within    & $M_{vir}$  & $M_{star}$ &$R_{vir}$ & $R_{\rm eff}$  \\
     & $R_{vir} (x 10^6)$ & ($10^{11}$\msun) &  ($10^{10}$\msun) & (kpc) &(kpc)\\
     (1)&(2)&(3)&(4)&(5)&(6)\\

 \hline
h239 & 17.2 & $9.13 $ & $4.50$ & 250  & 2.16 \\
h258 & 15.3 & $7.74$ &$4.46$  & 237 & 2.01 \\
h277 & 13.9 & $6.79 $  & $4.24$  & 227  & 2.41 \\
h285 & 16.6 & $8.82  $  & $4.56$ & 248 & 4.00\\
h603 & 21.8 & $3.43  $ &  $0.78$ & 181  & 3.77\\
\hline
\end{tabular}
\medskip 
Column 1:  Simulation name.  Column 2: Number of particles within $R_{vir}$. Column 3: Total mass within $R_{vir}$.    Column 4:  Stellar mass within $R_{vir}$.    Column 5: The virial radius $R_{vir}$.  Column 6: $R_{\rm eff}$ in the $V$ band measured for a face-on orientation.

\end{table}

\subsection{Theoretical Measurements of Kinematics and
  Shape}\label{sect:true}

One clear advantage of our galaxy models is that we can kinematically
select the bulge stars and measure their dispersion and ellipticities.
We follow the method of \citet{Abadi03} and begin by identifying the
stellar disk.  We orient the coordinate system so that the angular
momentum axis points along the $z$-axis and calculate $J_z$, the
angular momentum of each star in the $x-y$ plane.  We compare $J_z$ to
$J_{circ}$, the angular momentum the star would have if it were on a
circular orbit with the same energy.  We designate disk stars as
having $J_z / J_{circ} \geq 0.8$,.  To identify the spheroid, we
iteratively solve for the cutoff in $J_z / J_{circ} $ at which the
mean rotational velocity is zero.  This value differs for each galaxy
but tends to be around 0.5.  Using the entire matter distribution
(gas, stars, and dark matter), we calculate the total energy for each
particle in order to differentiate halo stars from the bulge.  Bulge
stars have higher binding energy than halo stars, and we use the
median value of the stars' total energy to distinguish the bulge from
the halo.

After kinematically identifying the bulge, we centre it in position
and velocity and determine the half-mass radius.  The stars within
this radius are those for which we measure $\sigma_{\rm tot}$;
however, we exclude stars within a radius of 0.3 kpc (see Section
\ref{sect:obs} and Figure \ref{fig:vprofiles}).  We calculate velocity
in the $x$, $y$, and $z$ directions for each star particle.  Summing
the variance of these quantities gives us the square of the ``true''
velocity dispersion, $\sigma_{\rm tot}$, measured directly from the
simulation.  We expect the three-dimensional dispersion to be a factor
of $\sqrt{3}$ smaller than a one-dimensional line-of-sight value for
an isotropic spheroid, and so we list the quantities $\sigma_{\rm
  tot}$ and $\sigma_{\rm tot}/\sqrt{3}$ in Table
\ref{table:sigmatable}.  We also measure the intrinsic shape by
calculating the moment of inertia tensor at the half mass radius.  Our
bulges are extremely realistic, and are consistent with the
measurements made by \citet{Christensen14a} which show that the bulges
obey the observed scaling relations relating surface brightness,
magnitude, and size.

\begin{table}
 \centering
 \caption{Comparison of Theoretical and Observed Measurements\label{table:sigmatable}}
  \begin{tabular}{@{}lccccc@{}}
    \hline
    Run     & Theoretical    & Theoretical & Median & Median  $\sigma_{\rm los}$ &Shape\\
    & $\sigma_{\rm tot}$  & $\sigma_{\rm los}$ & $\sigma_{\rm los}$   & (no rotation)  &($b/a$)\\
    (1)&(2)&(3)&(4)&(5)&(6)\\
    \hline
    h239 	& 191.7& 110.7	&  143.8 &  122.0   &0.95 \\
    h258 	& 193.8&111.9	&  147.6&   117.3    &0.60\\
    h277 	& 190.5& 110.0	&  141.6  &  113.7   &0.85\\
    h285 	& 217.2& 125.4	&  135.5  &  133.7   &0.88\\
    h603 	&  91.6& 52.9	&  63.1  &  52.2   &0.97\\
    \hline
\end{tabular}
\medskip 
Column 1: Simulation.  Column 2:  Theoretical velocity dispersion $\sigma_{\rm tot}$ in km s$^{-1}$.  Column 3: $\sigma_{\rm tot}/\sqrt{3}$ in km s$^{-1}$.  Column 4:  Median $\sigma_{\rm los}$  in km s$^{-1}$ as calculated by Equation \ref{eqn:sigma}.  Column 5:  Median $\sigma_{\rm los}$  in km s$^{-1}$ as calculated by Equation \ref{eqn:woo}.  Column 6:  Intrinsic shape.
\end{table}

\subsection{Synthetic Observations of Kinematics and
  Shape}\label{sect:obs}

We have developed a process which closely mimics the observational
method for determining $\sigma_{\rm los}$ from long-slit spectroscopy.
To capture the effect of orientation on the line-of-sight $\sigma_{\rm
  los}$ measurement, we centre each galaxy in position and velocity
space, and then rotate the galaxy along a series of angles, mimicking
various lines of sight.  We define two angles, $\theta$ and $\phi$,
which are measured in the polar and azimuthal directions, and rotate
the galaxy in uniform increments in cos($\theta$) and cos($\phi$) to
obtain a total of 1024 different lines of sight.

For each orientation, we estimate the surface brightness by treating
each star particle as a stellar population with a Kroupa IMF
\citep{Kroupa}.  Using the Starburst99 population synthesis code
\citep{Starburst99}, we input the age and metallicity of each particle
and receive magnitudes in several optical bands.  We do not include
dust obscuration, but we expect that neglecting dust will have minimal
effect (see Section \ref{sect:caveats} for more discussion).  By
converting the $V$ band magnitudes to luminosities, we sum over all
the stars to obtain the surface brightness.  We then fit a series of
concentric ellipses to the surface brightness and measure the
effective radius, $R_{\rm eff}$, as described in
\citet{BinneyMerrifield} equation 4.18.  We measure our bulge
quantities using all of the stars in the galaxy which fall along the
two-dimensional projection within an ellipse with semi-major axis
$R_{\rm eff}$.  We present the $R_{\rm eff}$ for a face-on orientation
for each galaxy in Column 6 of Table \ref{table:properties}.

For each rotation, we then align a slit along the major axis of the
rotated bulge.  The slit has a width of 50 pc, which corresponds to
1\arcsec\ at 10 Mpc, though varying this width has negligible effect
on our results.  We divide the slit into 50 bins and measure the mean
radial velocity ($v_{\rm los}$) and the dispersion ($\sigma$) for each.
Our results are also insensitive to the number of bins, as long as
this number is greater than $\sim 5$.

\begin{figure}
\begin{center}
\includegraphics[scale=0.4]{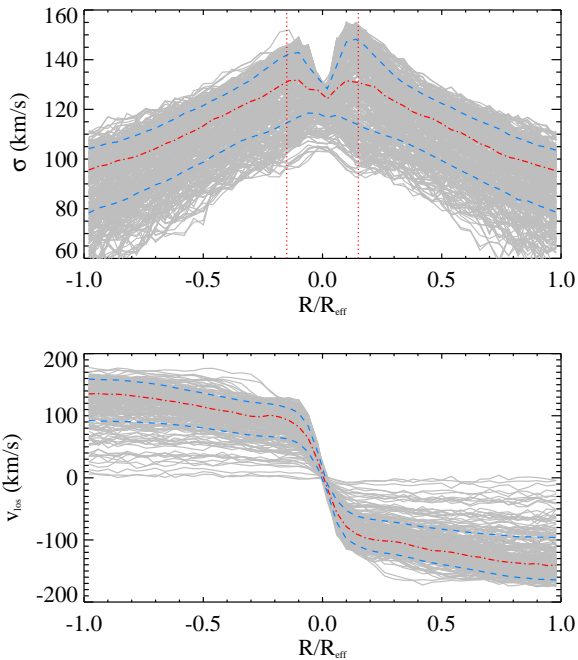}
\caption{ Profiles of $\sigma_{\rm los}$ (top) and $v_{\rm los}$ (bottom) from our simulated slit measurements for galaxy $h258$.  Grey lines represent measurements for the the full range of orientations, red dot-dashed lines are the medians, and blue dashed lines are the mean $\pm$ 68\%.  The vertical dotted lines in the top panel represent regions where resolution effects prevent an accurate measurement of $\sigma$, so we neglect that region of the slit.
\label{fig:vprofiles}
}

\end{center}
\end{figure}

 We then integrate along the slit, from $-R_{\rm eff}$ to $R_{\rm
   eff}$, using two methods.  Historically, studies have combined the
 velocity standard deviation $\sigma$ with the line of sight velocity
 $v_{\rm los}$ in accordance with the virial theorem, e.g. as in
 \citet{Gultekin09}:

\begin{equation}\label{eqn:sigma}
   \sigma_{\rm los}^2=\frac{\int{(\sigma(r)^2 + v_{los}(r)^2)I(r) dr}} {\int{I(r)dr}},
\end{equation}

\noindent
where $I(r)$ is the surface brightness.  However, a recent study by
\citet{Woo13} suggests that for systems with substantial rotation
(such as the disky galaxies we focus on here), the contribution of
$v_{\rm los}$ inflates the overall velocity dispersion.  In
an attempt to mitigate this bias, the authors instead suggest using
the more basic equation:

\begin{equation}\label{eqn:woo}
\sigma_{\rm los} = \frac{\int{\sigma (r) I(r) dr}}{\int{I(r)dr}}
\end{equation}

\noindent
for systems with a rotational component \citep[see also][]{Kang13}. It
is not clear, however, that ignoring the rotational or anisotropic
component of a virialized bulge would be appropriate to track a
theoretical link between the SMBH mass and the kinematics of the
bulge. For this reason, in this work we primarily focus on the first
method, but discuss how using Equation \ref{eqn:woo} affects our
results.  In Figure \ref{fig:vprofiles} we present our slit
measurements of $\sigma_{\rm los}$ (as calculated in Equation
\ref{eqn:sigma}) and $v_{\rm los}$ for every galaxy orientation (grey
lines) for the simulation $h258$.  We show the median and standard
deviation with red and blue curves, respectively.  While these
profiles are qualitatively similar to those presented in observational
papers, the central region of the velocity dispersion profile is not
well-represented due to the resolution limitations of our simulations.
We bracket the region with the central dip (which is about $\sim 15\%$
of $R_{\rm eff}$) with red dashed lines in Figure \ref{fig:vprofiles}.
This distance from peak to peak corresponds to a radius of $\sim 1.8$
softening lengths (or 0.3 kpc).  According to Figure
\ref{fig:vprofiles}, the stars in this region are not reliable for
kinematic study; we exclude this area from both our theoretical and
synthetic observation velocity dispersion measurements.  We have
tested our method by excluding ranges of 1, 2, and 3 times the
softening, and find that 1.8 is an appropriate factor to maximize
meaningful information while excluding that which is unreliable.  See
\S \ref{sect:caveats} for a discussion of resolution concerns.

We note that the bulges identified by our kinematic decomposition and
synthetic observations are not identical; each process selects the
bulge component based on different criteria.  At the moment it is not
clear whether either method is ``correct'' for measuring fundamental
scaling relations such as \ms.  We assert that observations of
$\sigma_{\rm los}$ may exhibit a large scatter due to galaxy
orientation, and that simulations can help explain the source of this
scatter, in part by determining the $\sigma$ from the
kinematically-selected bulge.

\section{Results}

\begin{figure*}
\includegraphics[width=6in]{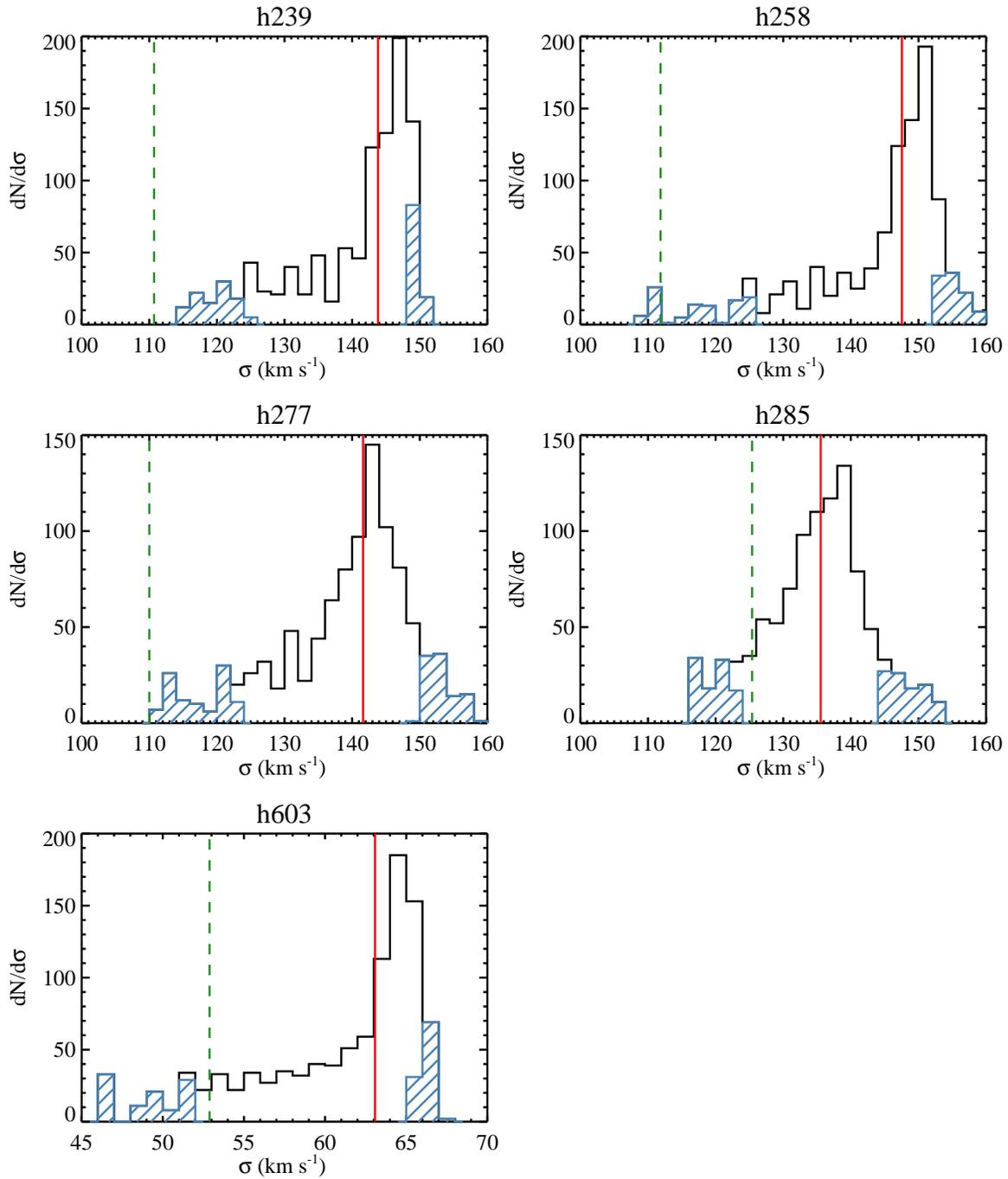}
\caption{Distribution of measurements of $\sigma_{\rm los}$ for every line of sight for all five galaxies.  The blue hatched regions are the lower and upper 10\% of the distributions, and the red vertical line is the median of each distribution.  The vertical green dashed line is the theoretical value of $\sigma_{\rm los}$ derived directly from the simulations (see \S \ref{sect:true}).
\label{fig:sigmadist}
}
\end{figure*}

The distribution of $\sigma_{\rm los}$ measurements (using Equation
\ref{eqn:sigma}) as a function of orientation of each galaxy is shown
in Figure \ref{fig:sigmadist}.  The red vertical lines are the medians
of each distribution, and the blue hatched regions are the highest and
lowest 10\% values.  The theoretically calculated line-of-sight
velocity dispersion is denoted by the vertical green dashed line for
each case.  The distribution is non-Gaussian for every galaxy, and is
skewed toward high $\sigma_{\rm los}$.  The fact that the distribution
is not Gaussian is disturbing; the effects of inclination cause the
{\em apparent} velocity dispersion to vary by several tens of km
s$^{-1}$, with a strong bias toward larger values.  The spread of
values is around 0.3 dex, consistent with what is observed in \ms
intrinsic scatter \citep{Gultekin09}.  Thus, these variations may be a
principal source of scatter in the low-mass end of the observed \ms
relation.  The scatter is larger than the stated observational
measurement errors, hinting that the wide spread in the low-mass end
of the \ms relation may be caused by evolutionary effects, or by an
underestimate of the measurement errors, or both \citep{Harris14}.
{\em We recommend that measurement errors for velocity dispersions of
  bulges in disk galaxies should never be estimated at less than 20 km
  s$^{-1}$, simply due to orientation.}

\begin{figure*}
\includegraphics[width=3in]{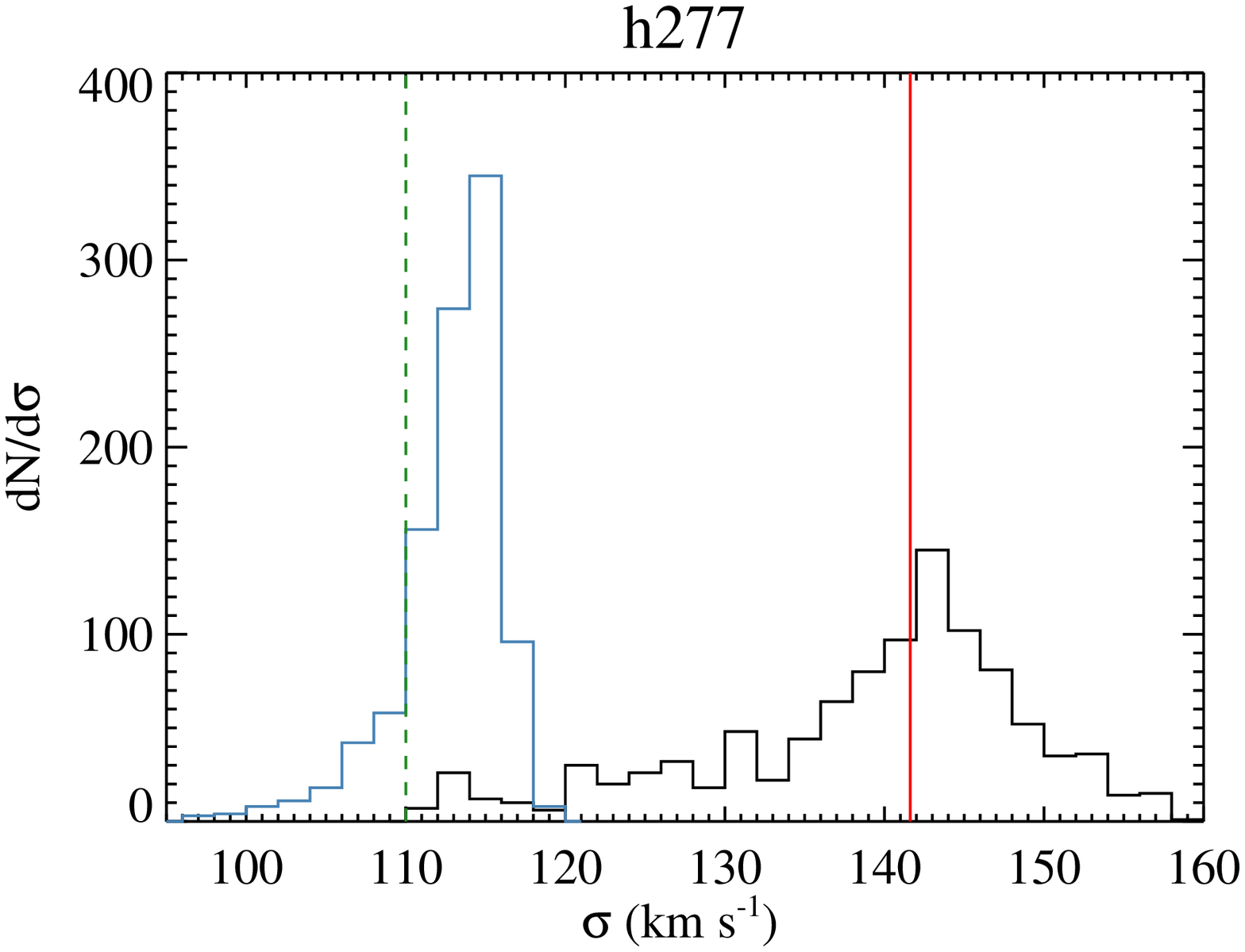}
\includegraphics[width=3in]{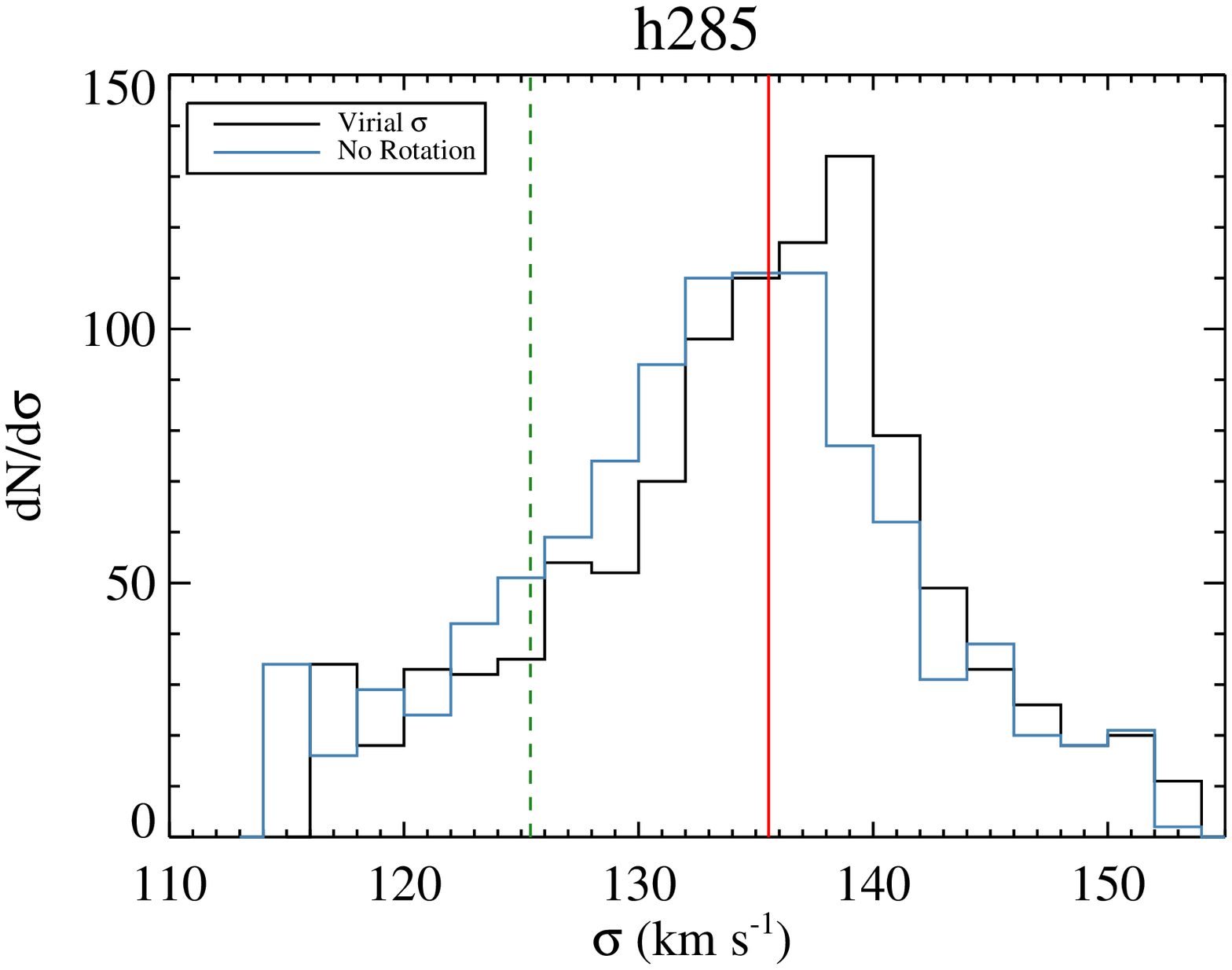}
\caption{{\em Left:} Distribution of $\sigma_{\rm los}$ calculated
  using Equation \ref{eqn:sigma} (black) and Equation \ref{eqn:woo}
  (blue) for simulation $h277$.  The solid red line indicates the
  median of the measured distribution, while the dashed green line is
  the theoretically measured line of sight value (see Section
  \ref{sect:true}).  {\em Right:} The same for simulation $h285$.  The
  median for $h277$ is shifted to lower values when the rotational
  component is removed.  However, $h285$ has no rotational component
  in the central region, so its $\sigma_{\rm los}$ profile is
  unchanged.
\label{fig:woo}
}
\end{figure*}

The use of Equation \ref{eqn:woo}, with the rotational velocity
component removed, has a marked difference for galaxies with a
noticeable rotational component.  In Table \ref{table:sigmatable} we
list the median values of $\sigma_{\rm los}$ for both observational
methods, and in Figure \ref{fig:woo}, we show the distribution of
$\sigma_{\rm los}$ calculated with both methods for two examples.  The
distribution without rotation (blue line) is shifted to lower values
for galaxy $h277$ (left panel), indicating that the inclusion of
rotational velocities contributes substantially to $\sigma_{\rm los}$.
The distribution is also far narrower, suggesting that the additional
rotational motions substantially broaden the range of possible
observed values.  On the other hand, galaxy $h285$ (right panel) does
not have substantial rotation in its central region, so the
distributions are indistinguishable.  Of our five galaxies, only
$h285$ lacks significant rotation in the central component; the other
four all show a {\em decrease of up to 25\% in their median
  $\sigma_{\rm los}$ values when Equation \ref{eqn:woo} is used}.
Overall, the method of Equation \ref{eqn:woo} is successful at
isolating purely dispersion-dominated motions, while Equation
\ref{eqn:sigma} represents the contribution of the full kinematic
system.  In terms of the \ms relation, it remains to be seen which
equation is a better metric to decipher how SMBHs and their host
bulges are interrelated (see \citet{Woo13} for more details).


In Table \ref{table:sigmatable} we compare the theoretical velocity
dispersion, $\sigma_{\rm tot}$, to those measured by synthetic
observations.  Comparing the median observed velocities for both
methods (with and without rotation) to the theoretical values, we see
that the simulation value is larger than the ``observed'' value, which
is in turn larger than the observed value neglecting rotation.  Our
estimates of $\sigma_{\rm tot}/\sqrt{3}$ fall within the extreme low
end of most of the line-of-sight measurements.  While this result
could be because the bulges are not perfectly spherical or isotropic,
the major factor is very likely a large population of non-bulge stars
contaminating the line of sight for the synthetic observations.  Since
it is impossible to isolate the bulge light from a two-dimensional
photometric projection, {\em this contamination factor will always be
  present.}

The observed measurement of $\sigma_{\rm los}$ excluding rotation is
characteristically lower (by $\sim 20$\%) than the traditional method,
which is understandable since there is no contamination by stars with
rotational motions.  The bulges of late-type galaxies have
non-negligible rotation, with the exception of $h285$\footnote{This
  galaxy is actually about to experience a merger, and is not in
  equilibrium, which may explain its lack of bulge rotation.}; in this
case the line-of-sight methods match each other.  Notably, the
measurement without rotation is very close to the one-dimensional
theoretical measurement.  Since both methods purposely exclude
rotational motions, it is reasonable that they should roughly agree.
Using Equations \ref{eqn:sigma} and \ref{eqn:woo} together give us an
idea of how rotation- vs dispersion-dominated a spheroid is; further
studies with such considerations may give us more clues to how SMBHs
grow and evolve with respect to the evolution of their hosts.



If $\sigma_{\rm los}$ measurements are larger for more inclined
systems, we expect to see a dependence of $\sigma_{\rm los}$ with the
inclination angle $\theta$.  Edge-on systems exhibit a large quantity
of disk stars along the line of sight, which inflate the observed
dispersion.  This geometrical argument has been made by
\citet{Brown13}, who studied the velocity structure of collisionless
simulations of disk galaxies using an integral field method.
Contamination by line-of-sight disk stars has been quantified by
\citet{Hartmann14} and \citet{Debattista13}, who suggest that a
highly-inclined system artificially boosts $\sigma_{\rm los}$ values
by 25\%; our results agree with this assessment.  However,
\citet{Graham09} used ellipticity as a proxy for inclination and found
no trend among the \ms residuals.  Our simulations actually do not
show clear trends of $\sigma_{\rm los}$ with ellipticity either; since
ellipticity changes with radius and may be affected by
non-axisymmetric shapes as well as inclination, we recommend that the
use of a kinematic estimate of inclination rather than one purely due
to shape.

\begin{figure*}
\includegraphics[width=6in]{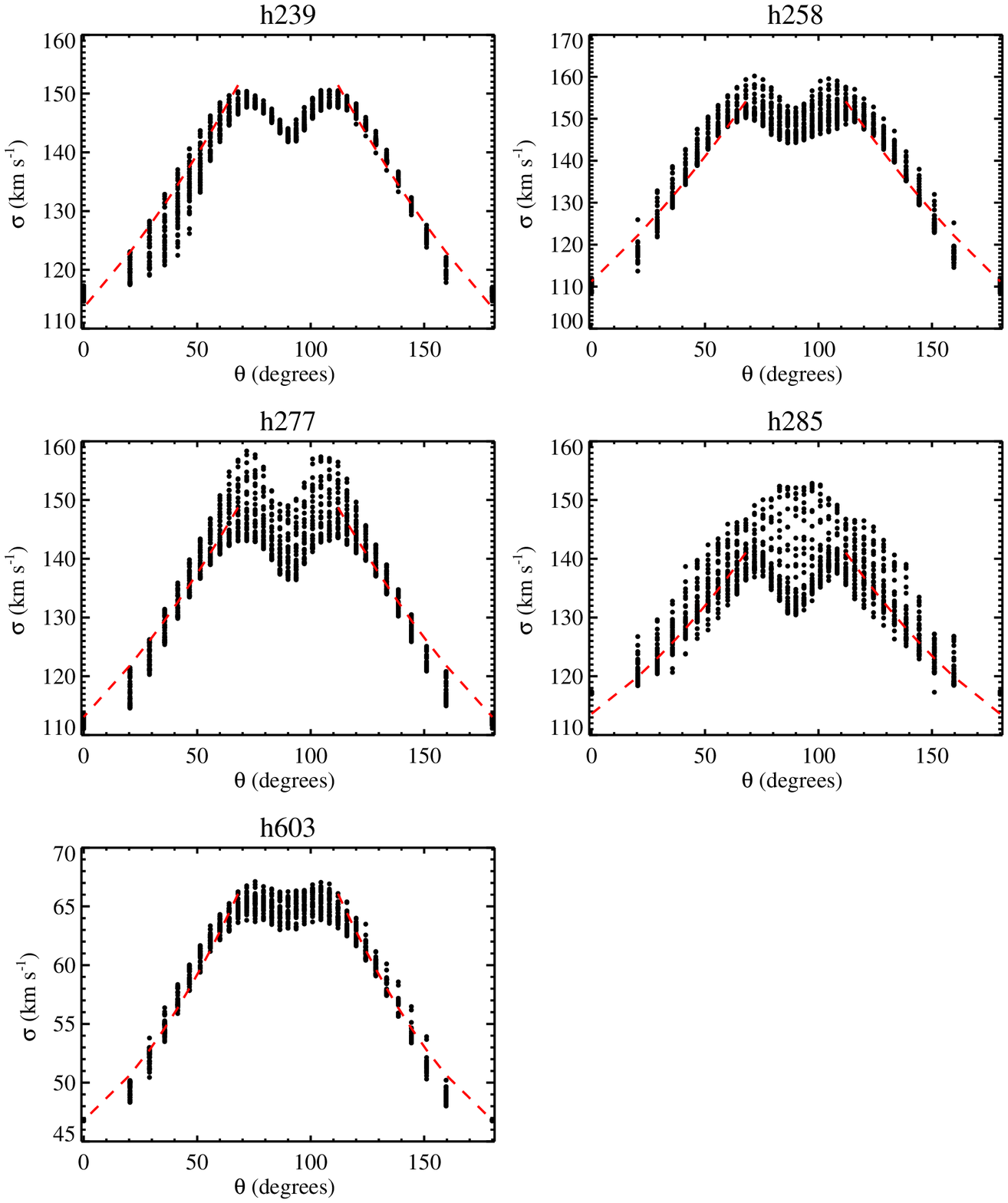}
\caption{  $\sigma_{\rm los}$ vs inclination $\theta$ for each line of sight for all five galaxies (black dots).  The red dashed line is the fit from Equation \ref{eqn:eureqa}.
\label{fig:thetasigma}
}
\end{figure*}

We find that the relationship between velocity dispersion and
orientation is somewhat straightforward -- in Figure
\ref{fig:thetasigma}, we plot velocity dispersion vs $\theta$ for all
1024 lines of sight for each galaxy (black points).  The points fall
in a fairly smooth curve, with some exceptions at edge-on orientations
where the scatter increases and the overall value of $\sigma_{\rm
  los}$ dips.  The increased scatter is due to the existence of
substructure and other anisotropies present in the galaxy.  The global
decrease in $\sigma_{\rm los}$ at high inclination is due to the fact
that the disk stars along the line of sight are moving both coherently
and with only a slight radial component, which decreases the overall
dispersion measurement.  Overall, face-on values of $\sigma_{\rm los}$
have the lowest scatter.

One of our goals with this work is to provide observers with an
inclination correction to more easily compare samples of galaxies and
to ascertain more realistic values for the intrinsic velocity
dispersion.  We employ Eureqa \citep{Schmidt09}, a machine learning
tool, to solve for a relation between $\theta$ and $\sigma_{\rm los}$.
We include two additional parameters, which are observationally
measurable: the circular velocity $v_{\rm rot}$ of the
galaxy\footnote{We measure $v_{\rm rot}$ by creating a synthetic HI
  emission line profile for an edge-on orientation and measuring
  $W_{20}/2$, the width of the line at 20\% of the peak.  See
  \citet{Governato09} for details.} in km s$^{-1}$, and the quantity
($v/\sigma$)$_{spec}$, measured from our simulated spectra at the
radius of influence (as in Figure \ref{fig:vprofiles}).  The galaxies
in our sample have varied ($v/\sigma$)$_{spec}$, reflecting the
different kinematics of each bulge, while $v_{\rm rot}$ is similar for
all but $h603$.  Adding these parameters allows us to include some
broader differentiating properties and create a universal model for
the relation between $\sigma_{\rm los}$ and inclination.  For this
fit, we neglect inclination angles $70 < \theta < 110$, due to the
increased scatter and drop in $\sigma_{\rm los}$ at very edge-on
orientations.  These orientations are not reliable for observationally
determined bulge measurements because the bulge is obscured by the
disk; in fact, such galaxies are commonly discarded from samples for
this reason.  We also add weight to the $\theta = 0$ values, because
these have the lowest scatter and will be the most useful for the
purposes of correcting to a universal orientation; it is vital that
our fit be excellent in this region.

Our equation is plotted as a red dashed line in Figure
\ref{fig:thetasigma} and is as follows:

\begin{equation}\label{eqn:eureqa}
\begin{aligned}
  \sigma_{\rm los} = 3.963v_{\rm rot} +
  0.003763 v_{\rm rot} \theta (\frac{v}{\sigma})_{spec}+ 0.001975\theta^2\\ -
  278 - 0.01003v_{\rm rot}^2 - 0.3187 \theta(\frac{v}{\sigma})_{spec}^2  \textrm{ km s}^{-1}
\end{aligned}
\end{equation} 

This relation between $\sigma_{\rm los}$ and $\theta$ allows us to
propose a correction for inclination effects.  The maximum error of
this fit for any line-of-sight measurement of $\sigma_{\rm los}$ is
10\%, and is generally less than 6\%.  We recommend observers correct
$\sigma_{\rm los}$ to a face-on value in order to compare samples of
galaxies with different orientations more carefully.  We caution the
use of measurements with inclinations larger than $70\,^{\circ}$, as
they are contaminated with a large number of non-bulge stars and
plagued by large scatter.

\section{Repercussions for the $M$--$\sigma$ Relation}

Thus far we have demonstrated that observational measurements of
$\sigma_{\rm los}$ may not be as reliable as previously thought.  This
revelation has many repercussions on galaxy dynamics and evolution.
In this section we focus on the effects on the observed \ms relation.
 
The low-mass end of the observed \ms relation has larger scatter than
the high-mass end \citep{Hu08,Graham11,Gultekin09,Gadotti09,Greene10}.
A common explanation for the scatter is simply hierarchical evolution;
as galaxies and black holes grow over time, they increase in mass
together and more tightly adhere to their scaling relations
\citep{Peng07,Jahnke10}.  Lower-mass galaxies in particular have
likely undergone fewer major mergers, and the above argument may not
even apply \citep{Kormendy11}.  SMBHs in low-mass galaxies may have
different growth mechanisms than their larger counterparts as well.
SMBH fueling in isolated disk galaxies may more likely be triggered by
secular, stochastic processes such as disk or bar instabilities
\citep{Cisternas11,Schawinski11,Kocevski12,Simmons13,Athanassoula13}
or by minor mergers \citep{Micic11,VanWassenhove12}.  Additionally,
mergers with other massive black holes may contribute substantially to
SMBH mass in low-mass galaxies \citep{Micic11}.  These process may not
cause the tight trends between SMBHs and larger-mass galaxy spheroids.
Evolutionarily speaking, the larger scatter for both $\sigma$ and
black hole mass for late-type galaxies is expected.

\begin{figure}
\includegraphics[scale=0.4]{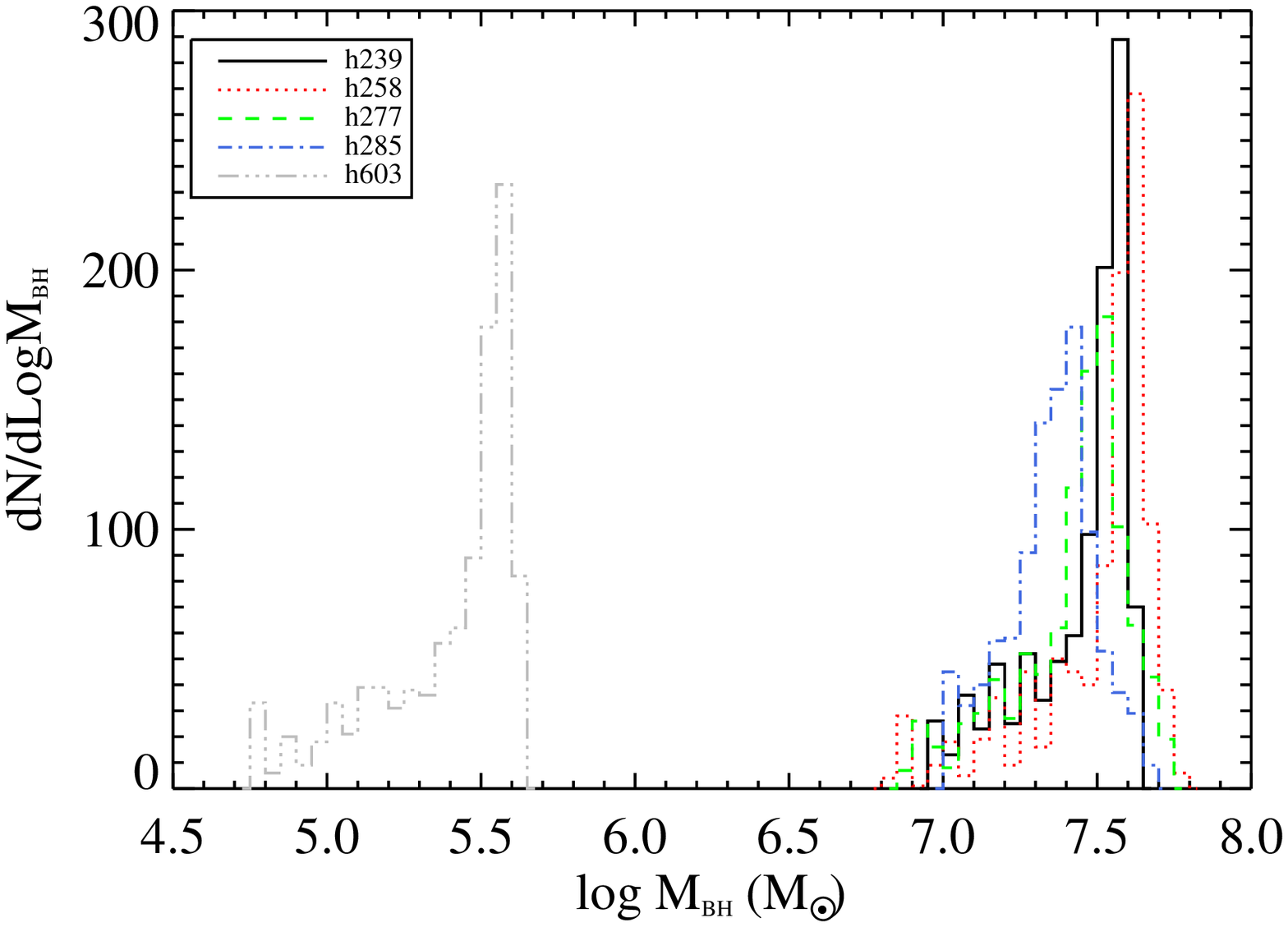}
\caption{ Distribution of black hole masses for each measurement of $\sigma_{\rm los}$, using the relation from \citet{McConnell13} to convert from $\sigma_{\rm los}$ to $M_{BH}$.  There is about an order of magnitude in the spread  of the estimated masses. 
\label{fig:bhmass}
}

\end{figure}

\begin{figure}
\includegraphics[scale=0.5]{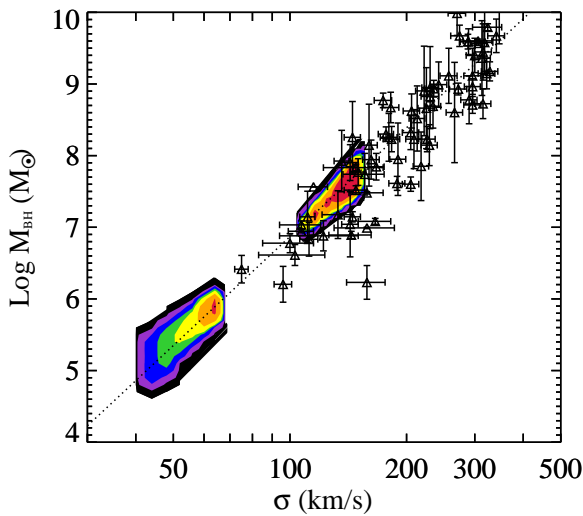}
\caption{Predicted scatter in the \ms relation based on the possible range of measured velocity dispersions.  We randomly draw $10^6$ values of $\sigma_{\rm los}$ for each galaxy, and assign a black hole mass based on the late-type relation from \citet{McConnell13}.  We include the intrinsic scatter in black hole mass as well as the uncertainty in their fit to create the distribution represented by the coloured contours (coloured by point density).  The black points and black dotted line represent the data and late-type fit to the data from \citet{McConnell13}.
\label{fig:msigma}
}

\end{figure}

However, the orientation effects presented in this paper may be able
to account for a substantial fraction of the scatter.  The
distributions of $\sigma_{\rm los}$ have a width of several tens of
km/s, and correspond to about 0.3 dex, which is approximately the
amount of scatter seen in the low-mass \ms relation.  We show how this
scatter translates to a scatter in estimated black hole mass in Figure
\ref{fig:bhmass}; using the relation from \citet{McConnell13}, we
input the values of $\sigma_{\rm los}$ for each line of sight to
obtain $M_{BH}$.  The values of $M_{BH}$ span about an order of
magnitude.  This wide scatter serves as a warning that estimates of
black hole masses from $\sigma_{\rm los}$ measurements may have much
larger errors than previously assumed (e.g. 0.33 dex in
\citet{Shankar04}) for late-type galaxies.  Conversely, theoretical
studies wishing to compare to the observed \ms relation must take care
to measure $\sigma_{\rm los}$ in a way that is consistent with
observational methods.

In Figure \ref{fig:msigma}, we depict the expected scatter in the
context of the \ms relation.  We randomly draw $10^6$ values of
$\sigma_{\rm los}$ for each galaxy from the measured distributions,
and assign a black hole mass based on the late-type relation from
\citet{McConnell13}.  We include the intrinsic scatter in black hole
mass as well as the uncertainty in their fit to create the
distribution represented by the coloured contours (coloured by point
density).  Overplotted on the figure are data points from
\citet{McConnell13}.  The scatter of the observational data is
comparable to, but somewhat larger than, the spread of the
simulations, indicating that evolutionary effects may have a role to
play as well, and that the scatter may not be due solely to
orientation effects.  \citet{Kormendy11} suggest that galaxies with
pseudobulges will exhibit larger scatter on the \ms relation, compared
to those with classical bulges.  Our single pseudobulge galaxy,
$h603$, does not demonstrate orientation-related behaviour that is
different from that of the more massive disk galaxies.  It is possible
that line-of-sight effects may not explain the increased scatter for
pseudobulges, but our sample is too small to make any robust
conclusions.  Regardless, it would be extremely informative to apply
our correction to data such as that in \citet{McConnell13}, and learn
what scatter remains after that due to inclination has been removed.

\section{Discussion and Conclusions}\label{sect:caveats}

While this work has focused on repercussions for the \ms relation, our
simulations do not include black hole physics such as accretion and
feedback.  We do not expect this exclusion to have a significant
effect; our galaxy sample is at a low enough mass that black hole
feedback effects do not dominate over other processes.  SMBH feedback
does affect star formation in the bulge region, and it is possible
that our velocity dispersion measurements are characteristically large
due to neglecting SMBH feedback quenching.  In fact, the bulge/disk
ratios in these simulated galaxies may already be larger than expected
compared to observations \citep{Christensen14a}; if SMBH feedback
reduces the size of the bulge relative to the disk, our results
concerning contamination from disk stars are likely strengthened.
Regardless, we do not expect our result of asymmetric $\sigma_{\rm
  los}$ distributions due to orientation to be changed in any way,
since the asymmetry is primarily caused by contamination from disk
stars and not by intrinsic bulge properties.

We also do not include the effects of internal dust extinction and
reddening when calculating surface brightness.  \citet{Stickley12}
employ a simple model for dust extinction and find that significant
dust presence can lead to a modest decrease ($\sim 13\%$) in the
measured value of $\sigma_{\rm los}$.  Since dust preferentially
effects edge-on orientations, it is possible that for these lines of
sight the observed $\sigma_{\rm los}$ would be lower.  We do not
expect any of our galaxies to be heavily obscured, however, and so our
results will not be affected strongly.

In \S \ref{sect:sims} we discuss eliminating the central region of
each simulated galaxy from our analysis for resolution reasons.  That
this step is necessary is unfortunate, because in observations the
highest signal-to-noise region is the centre, and there is no way to
compensate for its loss in a simulation with finite resolution.
Indeed, if we could include the central region, we expect that our
measurements of $\sigma_{los}$ using equations \ref{eqn:sigma} and
\ref{eqn:woo} may be brought closer into agreement, since this region
exhibits lower rotational velocities and higher dispersions.  However,
the main points of our results are not affected by this issue.  We
treat the theoretical measurements and synthetic observations in the
same manner, excluding the region from both, which assures we are
making valid comparisons.  In addition, the behaviour of $\sigma_{\rm
  los}$ out to $R_{\rm eff}$ is well-behaved outside of the excluded
region (Figure \ref{fig:vprofiles}), suggesting that the majority of
the data is of high quality.  We have also verified that increasing
the slit width (by up to a factor of 10) and adjusting the length of
the slit (by factors of a few in either direction) bring no
quantitative changes to our findings.  While the magnitude of our
$\sigma_{\rm los}$ measurements may be slightly underestimated because
we are missing the very peak of the central distribution, the
remainder of our results are still solid.

We caution the use of measurements of $\sigma_{\rm los}$ in late-type
galaxies to derive bulk galaxy properties.  In fact, any global
correlation that relies on $\sigma$, such as the Fundamental Plane,
will be biased.  The variation due to orientation alone is $\sim 20$
km s$^{-1}$, and the inability to eliminate disk stars from an
observational measurement introduces a contamination which
artificially increases $\sigma_{\rm los}$.  The method of
\citet{Woo13} may mitigate this effect somewhat; however, if a bulge
has a rotational component, the full kinematics will not be properly
accounted for.  We encourage the use of the relation of Equation
\ref{eqn:eureqa} to correct for orientation effects for inclinations
$\theta < 70\,^{\circ}$.

In summary, using state-of-the-art high resolution cosmological
simulations of disk galaxies, we quantify the effect of galaxy
orientation on the measurement of bulge velocity dispersion.  We
carefully designed our measurements to closely mimic observational
methods, and found that the value of $\sigma_{\rm los}$ is highly
dependent on viewing angle.  The distribution of $\sigma_{\rm los}$ is
asymmetric and skewed toward higher values, which correspond to more
inclined orientations.  The scatter in $\sigma_{\rm los}$ of $~\sim
0.3$ dex is approximately equal to that of the low-mass end of the \ms
relation, suggesting that orientation may substantially contribute to
the scatter.  Estimates of black hole masses using scaling relations
such as \ms must be taken with extreme caution in this range, as the
spread in $\sigma_{\rm los}$ corresponds to a 1.0 dex variation in
black hole mass.

\section*{Acknowledgments}

Simulations were run using computer resources and technical support
from NAS.  JMB and KHB acknowledge support from NSF CAREER award
AST-0847696.  FG acknowledges support from HST GO-1125 and NSF
AST-0908499. SL acknowledges the Michigan Society of Fellows for
financial support. CRC acknowledges support from NSF grants
AST-0908499 and AST-1009452. We are grateful to Jonathan Bird, Rololfo
Montez, Alister Graham, and Manodeep Sinha for helpful discussions,
and to the anonymous referee who provided extremely useful suggestions
which greatly improved the paper.


\bsp

\label{lastpage}

\end{document}